\documentclass{jaa}
\usepackage{natbib}
\usepackage{aas_macros}
\usepackage{graphicx}

\newcommand{\photu}{photon units}

\newcommand{\galex}{{\it GALEX}}

\begin{document}\sloppy

\title{Dust Scattered Radiation in the Galactic Poles }


\author{Jayant Murthy\textsuperscript{1,*}, R. C. Henry\textsuperscript{2} \and James Overduin\textsuperscript{3}}
\affilOne{\textsuperscript{1}Indian Institute of Astrophysics, Bengaluru 560 034, India.\\}
\affilTwo{\textsuperscript{2}Dept. of Physics and Astronomy, Johns Hopkins University, Baltimore, MD 21218, USA.\\}
\affilThree{\textsuperscript{3}Department of Physics, Astronomy and Geosciences, Towson University, Towson, MD 21252, USA\\}


\twocolumn[{

\maketitle

\corres{jmurthy@yahoo.com}

\msinfo{1 January 2015}{1 January 2015}

\begin{abstract}
We have modeled the diffuse background at the Galactic Poles in the far-ultraviolet (FUV: 1536 \AA) and the near-ultraviolet (NUV: 2316 \AA). The background is well-fit using a single-scattering dust model with an offset representing the extragalactic light plus any other contribution to the diffuse background. We have found a dust albedo of 0.35 -- 0.40 (FUV) and 0.11 -- 0.19 in the NGP ($b > 70^{\circ}$) and 0.46 -- 0.56 (FUV) and 0.31 -- 0.33 (NUV) in the SGP ($b < 70^{\circ}$. The differences in the albedo may reflect changes in the dust-to-gas ratio over the sky or in the dust distribution. We find offsets at zero-reddening of 273 -- 286  and 553 -- 581 photons cm$^{-2}$ s$^{-1}$ sr$^{-1}$ \AA$^{-1}$ in the FUV and NUV, respectively, in the NGP with similar values in the SGP.

\end{abstract}


}]


\doinum{12.3456/s78910-011-012-3}
\artcitid{\#\#\#\#}
\volnum{000}
\year{0000}
\pgrange{1--}
\setcounter{page}{1}
\lp{1}

\section{Introduction}

The diffuse ultraviolet sky was first observed by \citet{Hayakawa1969} and \citet{Lillie1969} who recognized that it was largely due to the scattering of starlight by interstellar dust grains \citep{Jura1979}. The diffuse Galactic light (DGL) was expected to be faintest at the Galactic poles and a number of rocket and satellite-based instruments observed the North Galactic Pole (NGP) in a search for extragalactic background light (EBL). These observations \citep{Anderson1979, Paresce1979, Paresce1980, Joubert1983, Jakobsen1984, Tennyson1988, Onaka1991, Feldman_hotgas1981, Hurwitz1991, Henry1993, murthy_model1995, Hamden2013, Murthy_dustmodel2016} typically found a surface brightness at the poles of about 200 -- 300 \photu\footnote{photons cm$^{-2}$ sr$^{-1}$ s$^{-1}$ sr$^{-1}$ \AA$^{-1}$} at about 1500 \AA. About half of this flux was attributed to the EBL \citep{Xu2005, Driver2016, Voyer2011, Gardner2000, Akshaya2018} with most of the remainder attributed to, as yet, unidentified sources \citep{Henry2015, kulkarni_fuv_2021} and residual dust-scaterring.

In this work, we will use observations in the far ultraviolet (FUV: 1536 \AA) and the near ultraviolet (NUV: 2316 \AA) made with the {\it Galaxy Evolution Explorer} (\galex) to model the dust-scattered component at high latitudes. Because the reddening is low at the Galactic poles, the scattering may be modeled using a single-scattering approximation and the UV scattering will be directly proportional to the reddening. This paper will focus on the model and the dust parameters with the expectation that we will map the extinction over the poles in a later work.

\section{Observations and Data Analysis}

The \galex\ spacecraft and its mission have been described by \citet{Martin2005} and \citet{Morrissey2007}. \galex\ included two photon-counting detectors: the far-ultraviolet (FUV: 1344 -- 1786 \AA\ with an effective wavelength of 1539 \AA) and the near-ultraviolet (NUV: 1771 -- 2831 \AA\ with an effective wavelength of 2316 \AA). The NUV instrument observed the sky for the entire length of the mission from 2003 June 7 until 2013 June 28 while the FUV detector failed permanently in 2009 May, with intermittent interruptions even before that date. The field of view (FOV) of the instrument was $1.25^{\circ}$ with a spatial resolution of 5 -- 10 arcseconds. 

Most of the sky was observed as part of the All-sky Imaging Survey (AIS) with a typical exposure time of about 100 seconds, with a few locations targeted for deeper observations of up to 100,000 seconds. These longer observations were comprised of a series of visits, each of less than 1000 seconds in length, taken during local (spacecraft) night and sometimes, but not always, taken in consecutive orbits. We \citep{Murthy2014apj, Murthy2014apss} extracted the diffuse background from the original \galex\ data after subtracting the foreground emission (airglow and zodiacal light) from each visit, available at a spatial resolution of $2'$ from https://archive.stsci.edu/prepds/uv-bkgd/.

\begin{figure*}
    \centering
    \includegraphics[width=3in]{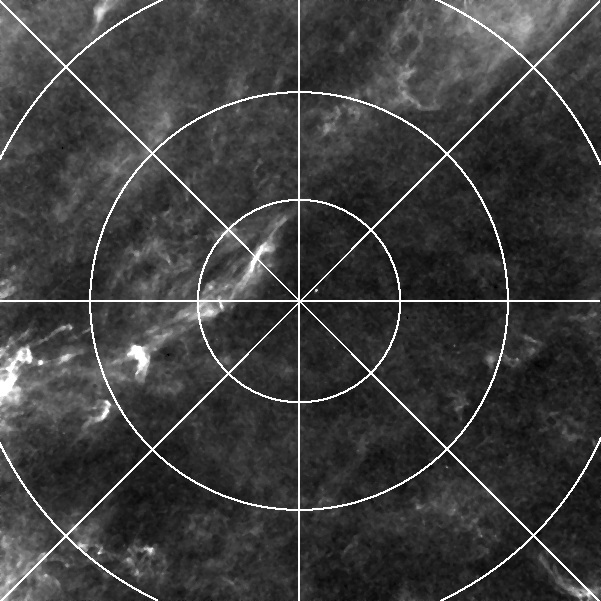}
    \includegraphics[width=3in]{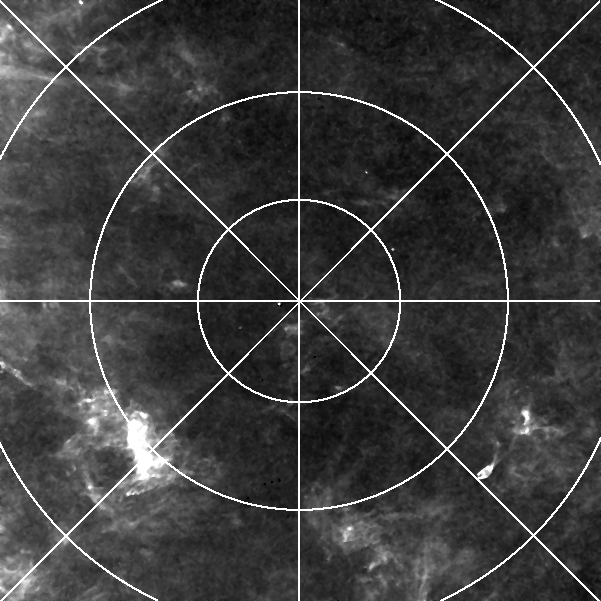}
    \caption{Planck E(B - V) map of the NGP (left) and SGP (right) scaled to a maximum E(B - V) of 0.1. Lines of longitude are plotted every 45 degrees with 0 degrees at the top of each plot. Lines of latitude are plotted every 10 degrees with the NGP at the center of the left image and the SGP at the center of the right image. We have restricted our analysis to $b > 70$ and $b < -70$, respectively.}
    \label{fig:ngpebv}
\end{figure*}

We have extracted the diffuse background from these files at both Galactic poles (SGP: b $< -70$; NGP: b $> 70$) from \citet{Murthy2014apj} with the Planck E(B - V) maps \citep{PlanckDust2016} shown in Fig. \ref{fig:ngpebv}. The mean reddening is on the order of 0.02 $\pm 0.005$ mag in both polar regions. In order to ensure that the dust-scattering is in the single-scattering regime, we have only used those areas with E(B - V) $<$ 0.05, excluding, for instance, the cirrus clouds observed by \citet{Markkanen1979}. Multiple scattering may become important at greater reddening and molecular hydrogen fluorescence \citep{Hurwitz1994} may contribute to the FUV emission.

\subsection{Time Variations}

\begin{figure}
    \centering
    \includegraphics[width=3in]{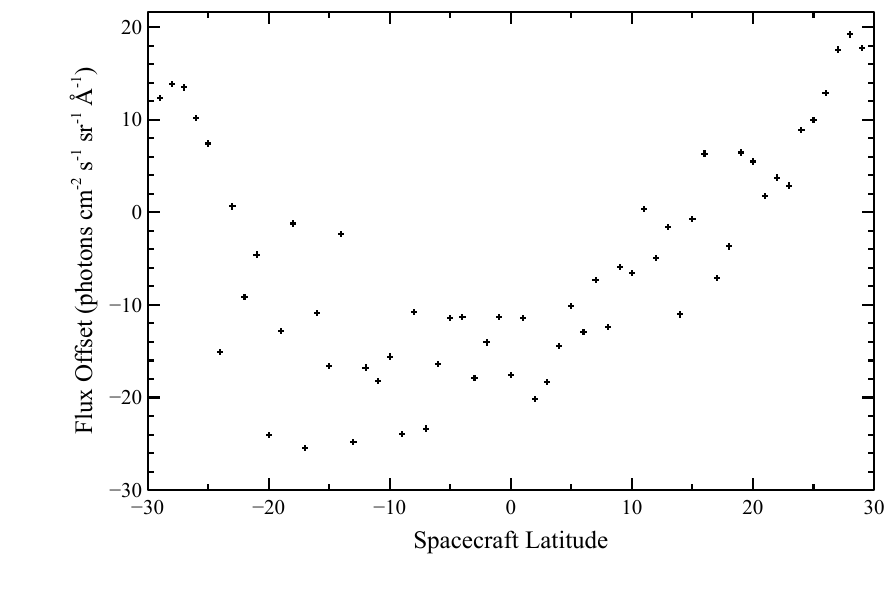}
    \caption{Variation of the NUV flux (difference from the mean) as a function of spacecraft latitude. The difference of about 50 \photu\ between the values at higher latitudes and the equatorial flux corresponds to an excess count rate of $1.2 \times 10^{-4}$ counts pixel$^{-1}$ s$^{-1}$. The corresponding rate in the FUV is $1.2 \times 10^{-5}$ counts pixel$^{-1}$ s$^{-1}$}
    \label{fig:latflux}
\end{figure}

\begin{figure}
    \centering
    \includegraphics[width=3in]{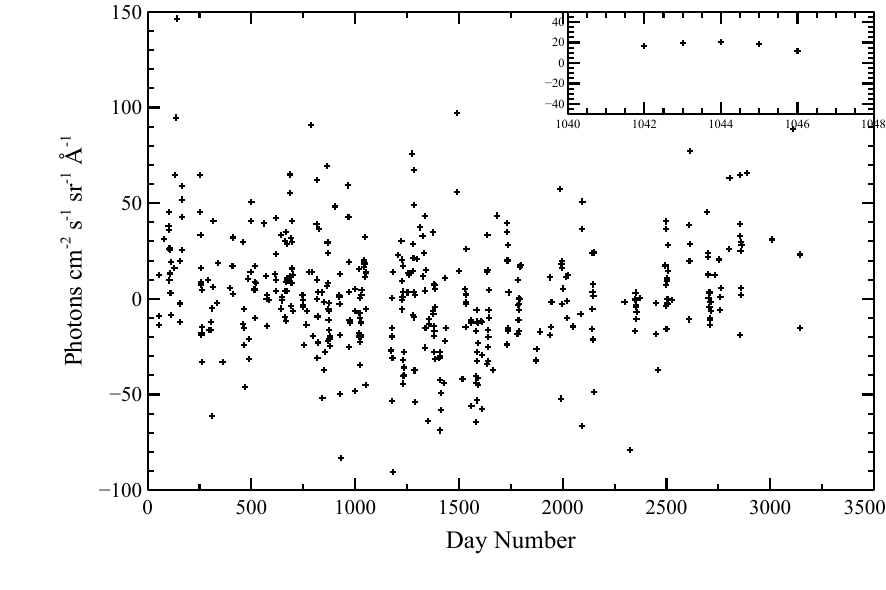}
    \caption{Variation of the NUV flux as a function of day number from 2003, June 10, with a short time period shown in the inset.}
    \label{fig:timeflux}
\end{figure}

\citet{Akshaya2018} noted variations of up to 200 \photu\ in individual visits of the Virgo region, although the deviation was much less in most observations \citep{Murthy2014apss}. We explore these deviations further using particularly deep observations in the NGP that had multiple visits spread over years finding that:
\begin{enumerate}
    \item The airglow increases with time from local midnight. We only included those observations that fell within two hours of orbital midnight, where the contribution from airglow is less than 5 \photu\ \citep{Murthy2014apss}.
    \item Observations within $60^{\circ}$ of the Moon showed off-axis scattering, particularly in the NUV. We excluded those observations.
    \item The emission in both bands increased at spacecraft latitudes greater than $20^{\circ}$ from the equator (Fig. \ref{fig:latflux}, suggesting a rise in the particle background at moderate latitudes. Interestingly, there is no dependence on distance from the South Atlantic Anomaly (SAA), even though the nearest observations were only $15^{\circ}$ away. We only included observations within 20$^{\circ}$ of the Earth's equator.
\end{enumerate}
Once these visits were excluded, we found no further variability in the diffuse background (Fig. \ref{fig:timeflux}) with a standard deviation of 20 \photu\ in the FUV and 30 \photu\ in the NUV for a 6 arcminute pixel.

\section{Modeling}

As mentioned above, the optical depth in the poles is small and the dust-scattered light is well-approximated by a single-scattering model. There are a relatively small number of stars contributing to the UV flux \citep{Henry_radiation_field1977} and we have used the Hipparcos catalog \citep{Perryman1997} as our source of stars, along with their positions, distances, and spectral types. We modeled each star using models from \citet{Castelli2004} and reddened the star using the 3-dimensional dust distribution of \citet{Green2019} with the  ''Milky Way'' extinction curve of \citet{Draine2003} to predict the ISRF (interstellar radiation field) at any location in space. The dust-scattered light in a given line-of-sight is the convolution of the ISRF with the local dust in a given bin, where the dust is distributed exponentially along the line of sight with a scale height of 125 pc \citep{Marshall2006} and a total reddening from \citet{PlanckDust2016}. We included the Local Bubble as a cavity of radius 50 pc around the Sun \citep{Welsh2010}.

\begin{figure}
    \centering
    \includegraphics[width=3in]{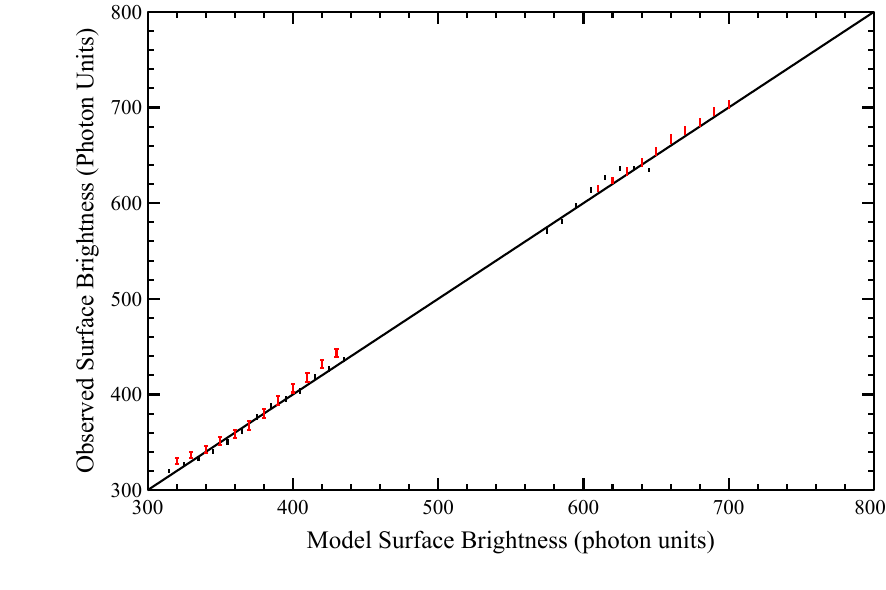}
    \caption{Model fit to the FUV and NUV observations at the NGP and the SGP (red), represented as $1 \sigma$ error bars with the FUV points in the lower left of the plot and the NUV in the upper right. Each point was calculated from the mean of the observations in that model bin, where the width of a bin is 10 \photu. The error bar is estimated from the photon noise. The solid line is the $x = y$ line.}
    \label{fig:model_data}
\end{figure}

\begin{table}
	\centering
	\caption{Best-Fit Parameters}
	\label{tab:opt_const}
	\begin{tabular}{llll}
		Band & $a$ & $g$ & Offset$^{a}$\\
		\hline
      NGP FUV & 0.35 -- 0.40 & 0.6 & 273 -- 286\\
      NGP NUV & 0.11 -- 0.19 & 0.5 & 553 -- 581\\
      SGP FUV & 0.46 -- 0.56 & 0.6 & 265 -- 289\\
      SGP NUV & 0.31 -- 0.33 & 0.5 & 580 -- 585\\
      \multicolumn{4}{c}{Predicted Values}\\
      FUV & 0.31$^{b}$ & 0.6$^{c}$ & 230 -- 290$^{d}$\\
      NUV & 0.36$^{b}$ & 0.5$^{c}$ & 480 -- 580$^{d}$\\
\hline
\multicolumn{4}{l}{Note: $g$ fixed to predicted values.}\\
\multicolumn{4}{l}{$^{a}$\photu}\\
\multicolumn{4}{l}{$^{b}$ \citet{Hensley2023}.}\\
\multicolumn{4}{l}{$^{c}$ \citet{Draine2003}.}\\
\multicolumn{4}{l}{$^{d}$ \citet{Akshaya2018}.}\\
\end{tabular}
\end{table}

\begin{figure}
    \centering
    \includegraphics[width=3in]{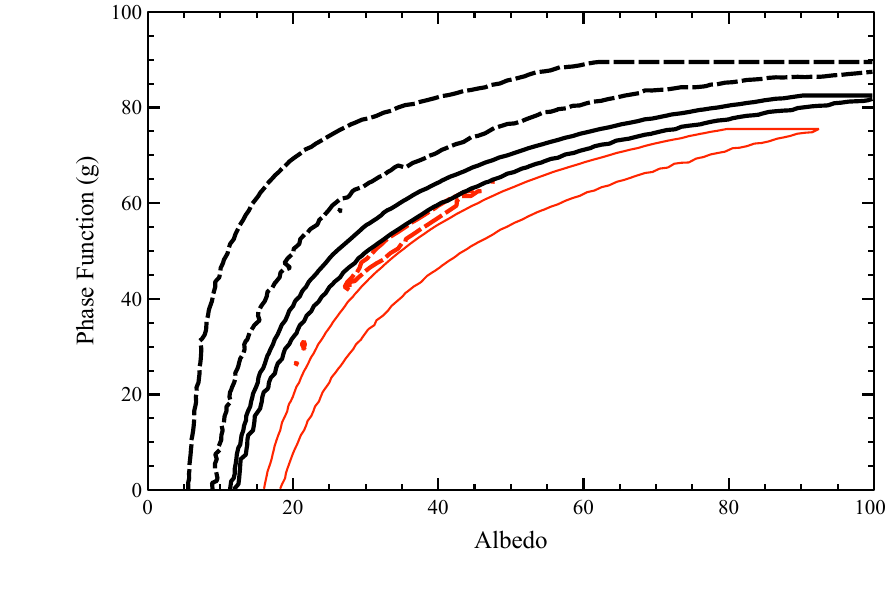}
    \caption{1 $\sigma$ contours for the optical constants in the NGP (black lines) and the SGP (red lines) with FUV parameters as solid lines and NUV as dashed lines.}
    \label{fig:ag}
\end{figure}
\begin{figure}
    \centering
    \includegraphics[width=3in]{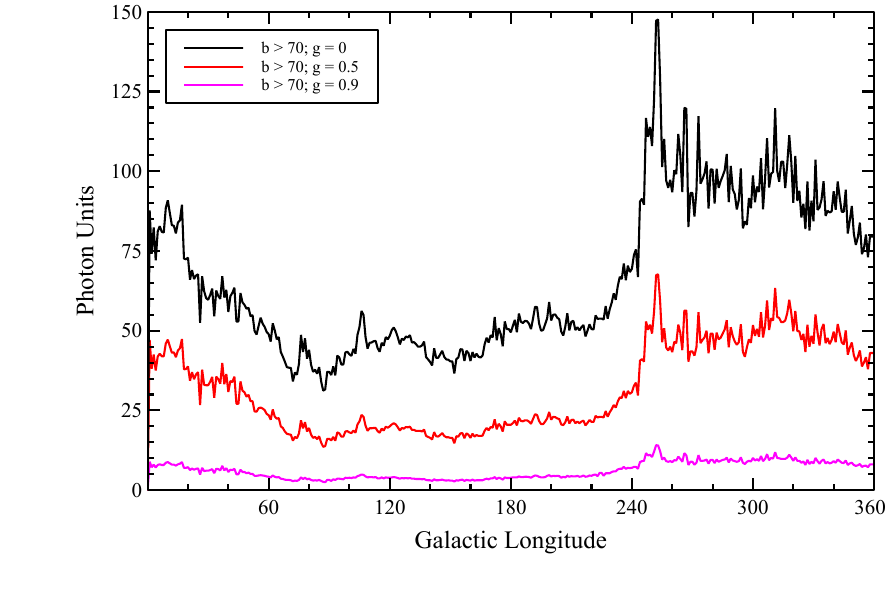}
    \caption{Scattered light in the FUV from dust at 60 pc for $b > 70$ for three values of $g$ and an albedo of 0.4.}
    \label{fig:long_model}
\end{figure}

\begin{figure}
    \centering
    \includegraphics[width=3in]{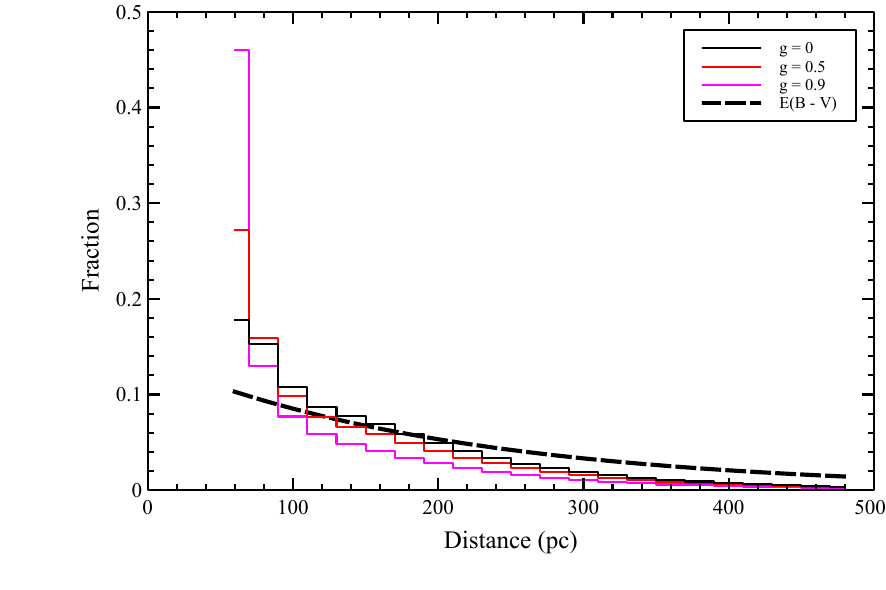}
    \caption{Relative contribution to the total diffuse radiation as a function of distance and $g$. The model has been averaged over the NGP (b $>$ 70). The relative dust distribution is plotted as the dashed line.}
    \label{fig:rad_model}
\end{figure}

We calculated models with different values of the optical constants ($a$ and $g$) and fit them to the observations in the Galactic poles, where we have averaged the individual observations into 10 \photu\ bins (Fig. \ref{fig:model_data}). The albedo ($a$) and the phase function asymmetry factor ($g$) are as defined by \citet{Henyey1941} and represent an empirical formulation of the dust scattering function. As such, our results are independent of the details of the dust models. The albedo and phase function are important constraints on dust models but have not yet been sufficiently well determined to be so used. The actual function may be more complicated \citep{Baes2022} but there is too much noise in our data to distinguish between scattering models and we will retain the Henyey-Greenstein function for consistency with previous results.

The error bars were calculated empirically from the data and were used to calculate the $\chi^{2}$ fits (Fig. \ref{fig:ag}), with the limits for $g = 0.6$ in the FUV and $g = 0.5$ in the NUV \citep{Draine2003} tabulated in Table \ref{tab:opt_const}. (The offsets between the model and the data represent the EBL and any additional component not associated with interstellar dust and were discussed by \citet{Akshaya2018}.) There is a trade-off between $a$ and $g$, where a higher albedo can compensate for the decreased radiation from more forward-scattering grains (Fig. \ref{fig:long_model}). We have found that the derived albedo is systematically higher in the SGP than in the NGP. This could be due to several factors, including a different dust-to-gas ratio or to modeling uncertainties in the dust distribution. Much of the scattered radiation comes from the dust closest to the Sun, especially for forward-scattering grains (Fig. \ref{fig:rad_model}), and relatively small uncertainties in the dust distribution may have a disproportionate effect on the observed radiation.

\vspace{-2em}
\section{Conclusion}

We have found that the diffuse radiation at the Galactic poles is well-fit by a relatively simple dust model with a single-scattering approximation. If we fix the phase function asymmetry factor ($g$) to 0.6 in the FUV, we find a dust albedo of 0.35 -- 0.40 in the NGP and 0.46 -- 0.56 in the SGP. Similarly, if we fix $g$ at 0.5 in the NUV, we find a dust albedo of  0.11 -- 0.19 in the NGP and 0.31 -- 0.33 in the SGP. 

There are many uncertainties in the modeling, although the data are surprisingly well-fit by our model, which may be indications that the dust-to-gas ratios differ across the sky \citep{Casandjian2022} or that the dust properties are different in the two poles. \citet{Henry2015} has pointed out the different behavior of the 100 $\mu$m thermal emission from the dust with the scattered component. In the next phase of our work, we will investigate these deviations and attempt to disentangle the different factors influencing the background observations in the poles.

\section*{Acknowledgements}
Part of this work was carried out while JM was a visitor in the Department of Physics, Astronomy and Geosciences at Towson University under the auspices of a Fundamental Physics Innovation Award funded by the Gordon and Betty Moore Foundation and administered by the American Physical Society.
We have used the Gnu Data Language (http://gnudatalanguage.sourceforge.net/index.php) and the Fawlty Language (https://www.flxpert.hu/fl/) for the analysis of these data. The data presented in this paper were obtained from the Mikulski Archive for Space Telescopes (MAST). STScI is operated by the Association of Universities for research in Astronomy, Inc., under the NASA contract NAS5-26555.  Support for MAST for non-HST data is provided by the NASA Office of Space Science via grant NNX09AF08G and by other grants and contracts. This research has made use of NASA's Astrophysics Data System.

\vspace{-1em}


\bibliography{murthy}

\end{document}